\documentclass[aps,preprint]{revtex4}%
\usepackage{amsfonts}
\usepackage{amsmath}
\usepackage{amssymb}
\usepackage{graphicx}%
\begin{document}
\title{Multiple Quantum NMR Dynamics in Pseudopure States }
\author{G. B. Furman}
\affiliation{Department of Physics, Ben Gurion University, Beer Sheva 84105, Israel}
\affiliation{Ohalo College, Qazrin, 12900, Israel}

\begin{abstract}
We investigate numerically the Multiple Quantum (MQ) NMR dynamics in systems
of nuclear spins 1/2 coupled by the dipole-dipole interactions in the case of
the pseudopure initial state. Simulations of the MQ NMR with the real
molecular structures such as six dipolar-coupled proton spins of a benzene,
hydroxyl proton chains in calcium hydroxyapatite and fluorine chains in
calcium fluorapatite open the way to experimental NMR testing of the obtained
results. It was found that multiple-spin correlations are created faster in
such experiments than in the usual MQ NMR experiments and can be used for the
investigation of many-spin dynamics of nuclear spins in solids

\end{abstract}
\startpage{1 }
\maketitle

\section{Introduction}

Method of multipe-quantum nuclear magnetic resonance (MQ NMR) in solids
\cite{baun1985} is a powerful tool for the investigations of structure and
spin dynamics in solids \cite{baum1986}. Significant progress has been made in
the simplification of ordinary NMR spectra \cite{warren1980}. Recently,
higher-order MQ NMR experiments have been utilized to address one of the
fundamental issues in quantum information processing, the preparation of the
pseudopure state \cite{J.-S. Lee2004}. \ A quantum computation includes as an
initial step preparation of the input state \cite{G. Benenti}. Conveniently,
the quantum algorithms start with a pure ground state where populations of all
states except the ground state are equal to zero. The realization of a pure
state in a real quantum system, such as a spin system, requires extremely low
temperatures and very high magnetic fields. To overcome this problem, a
so-called \textquotedblleft pseudopure\textquotedblright\ state was introduced
\cite{D. G. Cory,N. Gershenfeld}. The density matrix of the spin system in
this state can be partitioned into two parts. The first part of the matrix is
a scaled unit matrix, and the second part corresponds to a pure state. The
scaled unit matrix does not contribute to observables, and it is not changed
by unitary evolution transformations. Therefore, the behavior of a system in
the pseudopure state is exactly the same as it would be in the pure state.

Recently, it has been suggested \cite{furman2005,S. I. Doronin2007} to
consider the dipolar ordered state as the initial state for such experiments.
As a result of using the dipolar ordered initial condition, many-spin clusters
and correlations appear faster \cite{furman2005,S. I. Doronin2007} than in the
ordinary MQ NMR experiments \cite{baun1985} in solids and some peculiarities
of multiple-quantum (MQ) dynamics can be investigated with these experiments.

In the present article we consider MQ NMR dynamics when the initial condition
is determined by the pseudopure state. Our motivation of performance of this
work is defined first of all by that the many-spin correlations are created
faster in such experiments then in standard MQ NMR methods and can be used for
the investigation of many-spin dynamics of nuclear spins in solids. We
consider dynamics of MQ NMR in systems prepared in the pseudopure state and
calculate the resulting signal in MQ experiments. Note, that it is not
necessary to make any changes in the scheme of the standard experiment in
order to obtain non-zero signals of MQ coherences in the pseudopure state.
Computer simulations of such experiments for six spin ring and eight spin
linear chain are presented.

\section{MQ NMR with the initial pseudopure state}

We consider a system of nuclear spins ($s=1/2$) coupled by the dipole-dipole
interaction (DDI) in a strong external magnetic field $\overrightarrow{H_{0}}$
\ which is directed along the axis $z$. The secular part of the DDI
Hamiltonian \cite{goldman1970} has the following form
\begin{equation}
{\mathcal{H}}_{dz}=\sum_{j<k}D_{jk}\left[  I_{jz}I_{kz}-\frac{1}{4}\left(
I_{j}^{+}I_{k}^{-}+I_{j}^{-}I_{k}^{+}\right)  \right]  , \tag{1}%
\end{equation}
where $D_{jk}=\frac{\gamma^{2}\hslash}{r_{jk}^{3}}(1-3\cos^{2}\theta_{jk})$ is
the coupling constant between spins $j$ and $k$, $\gamma$ is the gyromagnetic
ratio, $r_{jk}$ is the distance between spins $j$ and $k$, $\theta_{jk}$ is
the angle between the internuclear vector $\overrightarrow{r_{jk}}$ and the
external magnetic field. $I_{j\alpha}$ is the projection of the angular spin
momentum operator on the axis $\alpha$ $(\alpha=x,y,z)$; $I_{j}^{+}$ and
$I_{j}^{-}$ are the raising and lowering operators of spin~$j$.

There are many pulse sequences exciting MQ coherences during the preparation
period. For a dipolar-coupled spin system, the multipulse sequence with an
eight-pulse cycle \cite{baun1985} is known to be very efficient. The basic
scheme of the standard MQ NMR experiments consists of four distinct periods of
time: preparation, evolution, mixing and detection \cite{baun1985}. In the
rotating reference frame \cite{goldman1970} the average Hamiltonian describing
the MQ dynamics at the preparation period can be presented as \cite{baun1985}
\begin{equation}
\mathcal{H}_{MQ}=\mathcal{H}^{(2)}+\mathcal{H}^{(-2)}, \tag{2}%
\end{equation}
where $H^{(\pm2)}=-\frac{1}{4}\sum_{j<k}D_{jk}I_{j}^{\pm}I_{k}^{\pm}$. \ Then
the evolution period without any pulses follows. The density matrix of the
spin system, $\rho(\tau)$, at the end of the preparation period is
\begin{equation}
\rho(\tau)=U(\tau)\rho(0)U^{+}(\tau), \tag{3}%
\end{equation}
where $U(\tau)=\exp(-i\tau(H^{(2)}+H^{(-2)}))$ is the evolution operator on
the preparation period and $\rho(0)$ is the initial density matrix of the
system. The transfer of the information about MQ coherences to the observable
magnetization occurs during the mixing period. Usually the thermodynamical
equilibrium density matrix is used as the initial one for MQ NMR experiments.
In the high temperature approximation the equilibrium density matrix takes a
form: $\rho(0)=I_{z}$ . Here we consider MQ NMR dynamics with the initial
pseudopure state when the density matrix can be described as:\qquad%
\begin{equation}
\rho(0)=\left\vert 1\right\rangle _{1}\otimes\left\vert 1\right\rangle
_{2}\otimes....\otimes\left\vert 1\right\rangle _{N}, \tag{4}%
\end{equation}
where $\left\vert 1\right\rangle _{k}$ represents a $k-$th spin that is up and
$N$ is the number of spins in the system. The method of creating the highly
polarized spin states (4) in clusters of coupled spins was described
previously \cite{J.-S. Lee2004,G. B. Furman}. It is based on filtering
multiple-quantum coherence of the highest order, followed by a time-reversal
period and partial saturation. It is convenient to expand the density matrix,
$\rho(\tau)$, at the end of the preparation period of the MQ NMR experiment as
\cite{feldman1997}
\begin{equation}
\rho(\tau)=\sum_{n}\rho_{n}(\tau), \tag{5}%
\end{equation}
where the term $\rho_{n}(\tau)$ is responsible for the MQ coherence of the
$n$-th order. One can find
\begin{equation}
e^{-i\delta I_{z}t}\rho_{n}(\tau)e^{i\delta I_{z}t}=e^{-in\delta t}\rho
_{n}(\tau). \tag{6}%
\end{equation}
On the mixing period the spin system is irradiated with sequence of the pulses
shifted on a $\pi/2-$phase regarding the pulses of the preparation period
\cite{baun1985}. As a result the average Hamiltonian describing evolution of
the spin system on the mixing period changes a sign on opposite to the sign of
the Hamiltonian (2), and the evolution operator, $U(\tau),$ is replaced by the
operator $U^{+}(\tau).$\ Starting with the initial conditions (4), the density
matrix, $\rho(t)$, after the three periods of the standard MQ NMR experiment
can be written as

\begin{equation}
\rho(t)=U^{+}(\tau)e^{-i\delta t_{1}I_{z}}\rho(\tau)e^{i\delta t_{1}I_{z}%
}U(\tau), \tag{7}%
\end{equation}
where $\rho(\tau)$\ is the density matrix at the end of the preparation period
according to Eq. (3) and $t=2\tau+t_{1}$, $\delta$ is the frequency offset on
the evolution period of the duration $t_{1}$ which is a result of applying the
time proportional phase incrementation (TPPI) method \cite{baun1985}. The
transfer of the information about MQ coherences to the observable
magnetization occurs during the mixing period. The last unitary transformation
in (7) with operator $U(\tau)$\ describes this period. The resulting signal
after the mixing period, the longitudinal magnetization, $M_{z}\left(
t\right)  $ , is
\begin{equation}
M_{z}\left(  t\right)  =Tr\left\{  \rho(t)I_{z}\right\}  \tag{8}%
\end{equation}
Using Eqs. (3) and (7) and the initial condition (4) it is convenient to
present the formula (8) for the longitudinal magnetization, $M_{z}\left(
t\right)  ,$ \ as follows%
\begin{equation}
M_{z}\left(  t\right)  =Tr\left\{  e^{-i\delta I_{z}t}\rho(\tau)e^{i\delta
I_{z}t}\rho_{MQ}(\tau)\right\}  \tag{9}%
\end{equation}
where
\begin{equation}
\rho_{_{MQ}}(\tau)=U(\tau)I_{z}U^{+}(\tau), \tag{10}%
\end{equation}
coincides with the density matrix at the end of the preparation period of the
\ standard MQ NMR experiment with the thermodynamical equilibrium density
matrix as the initial condition at high temperature approximation
\cite{baun1985}. The density matrix $\rho_{_{MQ}}(\tau)$ can be represented in
the following form \cite{feldman1997}
\begin{equation}
\rho_{_{MQ}}(\tau)=\sum_{n}\rho_{n}^{MQ}(\tau) \tag{11}%
\end{equation}
where the term $\rho_{n}^{MQ}(\tau)$ is responsible for the MQ coherence of
the $n$-th order and satisfies to the relationship of Eq. (6). By using Eqs.
(6), (7) and (9) one can rewrite the expression for the observable signal in
terms of the intensities of MQ coherences
\begin{equation}
M_{z}\left(  t\right)  =\sum_{n}e^{-in\delta t}J_{n}(\tau). \tag{12}%
\end{equation}
where $J_{n}(\tau)$ are the normalized intensities of MQ coherences are
\begin{equation}
J_{n}(\tau)=2Tr\left\{  \rho_{n}(\tau)\rho_{-n}^{MQ}(\tau)\right\}  /N.
\tag{13}%
\end{equation}
One can find from Eq. (13) that
\begin{equation}
J_{n}(\tau)=J_{-n}(\tau). \tag{14}%
\end{equation}
It is well known that in the usual MQ NMR experiments, the sum of the
intensities of all MQ coherences does not depend on time $\sum_{n}J_{n}%
(\tau)=1$. This is also right for the MQ NMR in the pseudopure state. \ 

\section{The numerical analysis of the time evolution of MQ coherences in the
pseudopure state}

We restrict ourselves to numerical simulations of MQ NMR dynamics of
one-dimensional systems. For example, quasi-one-dimensional six
dipolar-coupled proton spins of a benzene molecule \cite{J.-S. Lee2004} and
hydroxyl proton chains in calcium hydroxyapatite $Ca_{5}(OH)(PO_{4})_{3}$
\cite{Cho1996} and fluorine chains in calcium fluorapatite $Ca_{5}%
F(PO_{4})_{3}$ \cite{Cho1996} are suitable objects to study MQ dynamics in
pseudopure states. The numerical calculations were performed for MQ NMR
dynamics of linear chains and rings of 6 and 8 spins. The DDI coupling
constant of the nearest neighbors is chosen to be $D_{j,j+1}=D=1s^{-1}$. Then
the coupling constants of spins $j$ and $k$ are $D\left[  \frac{\sin\left(
\pi/N\right)  }{\sin\left(  \pi\left(  j-k\right)  /N\right)  }\right]  ^{3}$
for ring and $D/|j-k|^{3}$ for chain, respectively. In order to compare the
results of the numerical simulations with the analogous ones for the ordinary
MQ NMR dynamics we introduce the normalized intensities of MQ coherences for
the $n$ order, $J_{n}^{ord}(\tau)$ \cite{baun1985}
\begin{equation}
J_{n}^{ord}(\tau)=Tr\left\{  \rho_{n}^{MQ}(\tau)\rho_{-n}^{MQ}(\tau)\right\}
/Tr\left\{  I_{z}^{2}\right\}  . \tag{15}%
\end{equation}
The dependences of the intensities of MQ coherences on the dimensionless time,
$t=D\tau$ , of the preparation period in spin ring containing six spins are
presented in Fig. 1 for the both initial states, equilibrium (a) and
pseudopure (b). One can compare the intensities of MQ coherences in the
suggested experiment with the standard ones. It is evident that the suggested
method can be considered as a useful addition to the standard MQ NMR method.
Fig. 1 demonstrates that the suggested method yields the intensities of MQ
coherences of the fourth and six orders which several times higher than the
analogous coherence in the standard MQ NMR experiment. It is clear from the
insets of Fig. 1 that MQ coherence of the fourth and sixth orders, obtained
with the initial condition (4), appear little earlier than in the usual MQ NMR
in a ring of six spins. For example, or the pseudopure state at time $D\tau=2$
the MQ intensity of the fourth order coherence is $0.05$ (red dash line) while
for the usually used initial condition up to $D\tau=3$\ the intensity $J_{4}$
is closed to zero (black solid line). A tendency of the faster growth of MQ
coherences of high orders takes place also for the linear chain containing
eight spins (Fig. 2). This peculiarity is connected with the initial
pseudopure state (4). As a result, many-spin clusters and correlations
connected with MQ coherences appear faster than in the standard MQ NMR with
the initial condition, $\rho(0)=I_{z}$. The numerical calculations confirm
also the results obtained in the previous section. In particular, the growth
of MQ coherences occurs in accordance with Eq. (14).

One can see from Fig. 2 that the observable intensities of the fourth and
sixth orders in the linear chain containing eight spins can be negative in
contrast to the ordinary MQ NMR experiments at high temperatures
\cite{baun1985,feldman1997,doronin2000}. At the same time, it was shown
\cite{feldman2002} that the intensities of MQ coherences can be negative in
the standard NMR experiments at low temperatures and in the dipolar ordered
state at high temperatures. In fact, in MQ NMR experiments the observable
quantity is the longitudinal magnetization modulated by rf pulses. The
distinct frequency components of the magnetization can have an arbitrary sign.

\section{Conclusions}

The MQ NMR method for the detection of MQ coherences starting from the
pseudopure state is proposed. Investigations of MQ NMR dynamics in the
pseudopure states can be considered as a supplementary method which
complements the usual NMR in order to study structures and dynamical processes
in solids. Many-spin clusters and many-spin correlations are created faster in
such experiments than in the usual MQ NMR with the initial equilibrium
condition without any correlation between the spins. In this paper we focused
on simple one-dimensional examples but the physical picture obtained here is
not limited to perform simulations and experiments in two-dimensional and
three dimensional systems and open new ways for the study of many-spin systems.

\subsection*{Acknowledgements}

Author would like to thank Ben Gurion University of the Negev for supporting
this work.

\bigskip

Caption figures

Fig. 1 (Color online) The time dependence of the intensities of the MQ
coherences in a ring of six spins coupled by the DDI: (a) termal equilibrium
initial state and (b) the pseudopure initial state (4). Black solid line:
intensities of the zeroth order coherence $J_{0}$ . Red dash line: intensities
of the second order coherence $J_{2}$ . Green dot line: intensities of the
fourth order coherence $J_{4}$ . Blue dash-dot line: intensities of the sixth
order coherence $J_{6}$ . The insets show that the MQ coherences of fourth (a)
and sixth (b) orders in the pseudopure state (red dash line) appear little
earlier than in the usual MQ NMR (black solid line).

\bigskip

Fig. 2 (Color online) The time dependence of the intensities of the MQ
coherences in a linear chain of eight spins coupled by the DDI: (a) termal
equilibrium initial state and (b) the pseudopure initial state (4). Black
solid line : intensities of the zeroth order coherence $J_{0}$ . Red dash
line: intensities of the second order coherence $J_{2}$ . Green dot line:
intensities of the fourth order coherence $J_{4}$ . Blue dash-dot line:
intensities of the sixth order coherence $J_{6}$ . The insets show that the MQ
coherences of fourth (a) and sixth (b) orders in the pseudopure state (red
dash line) appear little earlier than in the usual MQ NMR (black solid line).

\bigskip
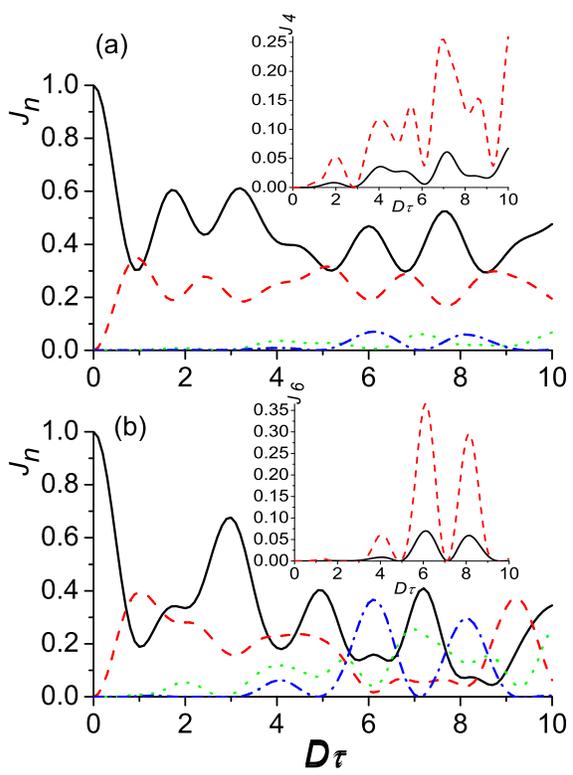
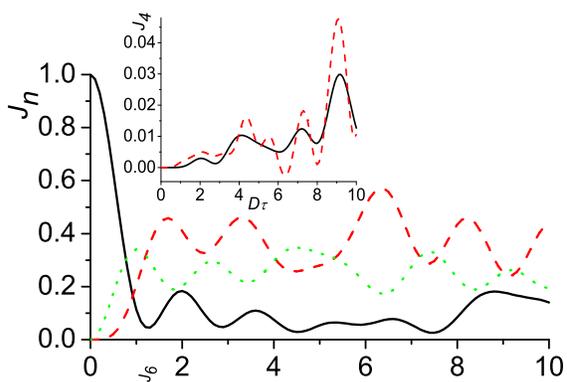
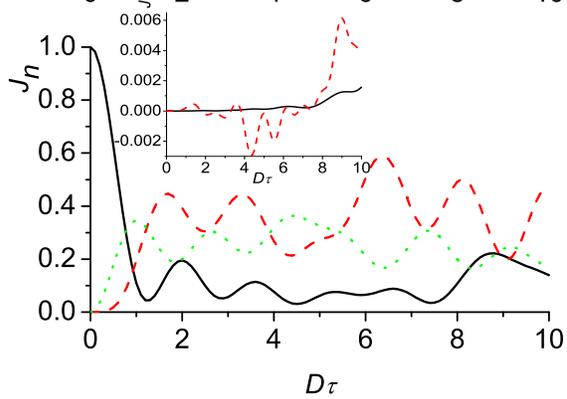
\end{document}